\documentclass[
  a4paper,
  twocolumn, %appr. journal style! You can also choose 'preprint' or 'twocolumn'
  prl, 10pt,
  superscriptaddress,
  amsfonts,amsmath,amssymb,
  longbibliography
]{revtex4-2}
\usepackage[
  colorlinks,
  citecolor=blue,
  linkcolor=blue,
  urlcolor=blue
]{hyperref}
\usepackage{graphicx,epstopdf,bm}
\usepackage{orcidlink}
\usepackage{braket}
\usepackage[capitalize]{cleveref}
\usepackage{xcolor}
\usepackage[all]{hypcap}
\usepackage{multirow}

\usepackage{lipsum}\setlipsum{%
  par-before = \begingroup\color{lightgray},
  par-after = \endgroup
}
\crefname{section}{Sec.}{Secs.}
\Crefname{section}{Section}{Sections}
\crefname{appendix}{App.}{Appces.}
\Crefname{appendix}{Appendix}{Appendices}

\DeclareMathOperator{\Tr}{Tr}

\newcommand{\op}[1]{\hat{#1}}

\newcommand{\figref}[2][]{Fig.~\hyperref[#2]{\ref*{#2}#1}}
\newcommand{\figureref}[2][]{Figure~\hyperref[#2]{\ref*{#2}#1}}

\newcommand{\sqrtiswap}{$\sqrt{i\mathrm{SWAP}}$}

\newcommand{\emref}{\hyperref[sec:end-matter]{End Matter}}
\begin{document}

%\title{Optimizing the Perfect Entangler Spectrum --- Crosstalk Mitigation with Optimal Control}
\title{Mitigating Dynamic Crosstalk with Optimal Control}
%Minimal Pulse Changes \\ Using the Perfect Entangler Spectrum}
%Optimization of the Perfect Entangler Spectrum} %Alternativtitel

%%% Authors %%%%%%%%%%%%%%%%%%%%%%%%%%%%%%%%%%%%%%%%%%%%%%%%%%%%%%%%%%%%%%%%%%%%

\author{Matthias G. Krauss}
\author{Luise C. Butzke}
\author{Christiane P. Koch\orcidlink{0000-0001-6285-5766}}
\affiliation{Freie Universit\"{a}t Berlin, Dahlem Center for Complex Quantum Systems and Fachbereich Physik, Arnimallee 14, 14195 Berlin, Germany
}

%###############################################################################

\begin{abstract}
The prevalence of quantum crosstalk is an important barrier to scaling frequency-addressable qubit architectures, with dynamic crosstalk  being particularly difficult to detect and suppress. This form of crosstalk refers to unintended interactions driven by the gate control fields themselves. Here, we minimize dynamic crosstalk using quantum optimal control based on the perfect entangler spectrum, where spectral peaks signal unwanted entanglement with spectator qubits. Focusing on parametric gates in tunable coupler systems, we derive pulse shapes that eliminate dynamic crosstalk. %While suppressing crosstalk due to spectral overlap of different qubits requires highly complex drives, 
Remarkably, only minimal pulse modifications are required to mitigate the form of crosstalk that is otherwise most difficult to predict.
%arising from drives resonant with single-qubit transitions. %The dynamics underlying these resonances is inherently exploited to find high-fidelity solutions. In contrast, 
%  Because this latter type of crosstalk is directly inferable from the Hamiltonian, it should be addressed by hardware design rather than pulse shaping.
  The ability to suppress dynamic crosstalk via the perfect entangler spectrum establishes a generalizable control principle for eliminating unwanted interactions in quantum hardware.
\end{abstract}

%###############################################################################

\maketitle
%\tableofcontents

%###############################################################################

%==================================================
\paragraph{Introduction}
%==================================================

The realization of large-scale quantum computation~\cite{Nielsen10} hinges on the high-fidelity control of complex many-body systems. In contemporary quantum hardware, however, system-environment interactions and crosstalk are still substantial and limit both coherence times and gate fidelities. As a form of coherent error arising from unintended interactions, quantum crosstalk~\cite{SarovarQ20} is particularly prevalent in architectures utilizing frequency-addressable control. It constitutes a fundamental barrier to the implementation of both quantum error correction~\cite{ParradoRodriguez2021crosstalk,chen_exponential_2021} and quantum algorithms~\cite{Ding20,Murali2020}.
%, where unique transition frequencies are assigned to individual qubits,  allowing for selective manipulation via tuned electromagnetic pulses. 

Attempts to mitigate crosstalk have 
%focused on static crosstalk, in particular residual $ZZ$ interactions, %and superconducting qubit architectures, and 
addressed various layers of the quantum computing stack, from hardware design~\cite{MundadaPRAppl19,KuPRL2020,KandalaPRL21,ZhaoPhysRevApplied23} to qubit control~\cite{SungPRX21,ParradoRodriguez2021crosstalk,TripathiPRAppl22,ShirizlyPRL2025,NiuQST2024,EvertPRAppl25}, error correction~\cite{zhou2025surfacecodeerrorcorrection}, and circuit compilation~\cite{Ding20,Murali2020,PerrinPRR24,Wagner25}. 
At the hardware layer, static interactions can be suppressed  by %suitable design of the two-qubit level spacings, either by 
coupling qubits with opposite sign or leveraging 
%the interference of multiple coupling paths~\cite{MundadaPRAppl19,KandalaPRL21}.
interference~\cite{MundadaPRAppl19,KandalaPRL21}. 
At the qubit control and gate implementation layer, pulse shaping~\cite{SungPRX21} and dynamical decoupling~\cite{ParradoRodriguez2021crosstalk,TripathiPRAppl22,ShirizlyPRL2025,NiuQST2024,EvertPRAppl25} mitigate crosstalk caused by static interactions. Dynamical decoupling also addresses single-qubit decoherence~\cite{ParradoRodriguez2021crosstalk,TripathiPRAppl22,ShirizlyPRL2025,NiuQST2024,EvertPRAppl25}, but comes at the expense of longer circuits~\cite{BrownPRXQ2025}. At the instruction layer, crosstalk can be minimized by the choice of gates~\cite{GanzhornPRR20},
%randomized~\cite{PerrinPRR24},  frequency-aware~\cite{Ding20} or symmetry-aware~\cite{VezvaeePRXQ2025} 
compilation~\cite{Ding20,PerrinPRR24,VezvaeePRXQ2025}, % as well as 
scheduling~\cite{Murali2020} and routing~\cite{Wagner25}.
%%$
In contrast to errors due to always-on static interactions, dynamic crosstalk arises during the gate implementation from the control fields themselves~\cite{ZhaoPhysRevApplied23}. For single-qubit gates, dynamic crosstalk can be suppressed by designing control pulses that satisfy analytical conditions derived from a crosstalk error model~\cite{ZhouPRL23} or by exploiting the inherent freedom of Bloch-sphere paths in unconventional geometric computing~\cite{HongPRAppl24}. 
Extension to entangling gates is possible for two-level systems~\cite{LiangPRApp2024}, but such a framework neglects the primary mechanisms of dynamic crosstalk that involve higher-energy transitions~\cite{MalekakhlaghPRA2020}. 
While static interactions and drive-induced crosstalk during local operations can be accounted for, multi-qubit coherent errors are notoriously difficult to characterize~\cite{KrinnerPRA20}, and their suppression is an outstanding challenge.

In this Letter, we show how to minimize dynamic crosstalk in multi-qubit operations.
To this end, we combine the perfect entangler (PE) spectrum~\cite{Krauss25} with quantum optimal control theory~\cite{KochEPJQT22} to derive pulses that enhance the gate fidelity of typical entangling operations by three orders of magnitude. 
By utilizing an optimization functional based on the perfect entangler spectrum, we bypass explicit characterization of crosstalk, as the design process inherently captures and suppresses unwanted interactions with spectator qubits. We exemplify our approach using superconducting qubits with tunable couplers, demonstrating that minimal pulse modifications are sufficient to eliminate those crosstalk errors that are most difficult to diagnose.

%==================================================
\paragraph{Optimal control with the perfect entangler spectrum}
%==================================================

Quantum optimal control constitutes a versatile toolkit for designing pulse shapes of external drives that implement a desired quantum dynamics~\cite{KochEPJQT22}. The task is cast as an optimization problem: a cost functional is defined to quantify performance, then maximized or minimized to determine the field profile.
The most direct approach to realizing crosstalk-free gates is to minimize the total gate error within a model encompassing both target and spectator qubits. However, such a "black-box" minimization obscures the physical mechanisms by which crosstalk is mitigated. A more systematic alternative is provided by the PE spectrum~\cite{Krauss25}, a diagnostic tool that identifies dynamic crosstalk in universal two-qubit gates. As a function of the base frequency of a third spectator qubit, the PE spectrum quantifies parasitic entanglement between that spectator and the computational qubits. Adopting the PE spectrum as the cost functional directly targets crosstalk suppression as the primary control objective.

Formally, the PE spectrum is defined as~\cite{Krauss25}
\begin{equation}\label{eq:J}
    \mathcal{J}_\mathrm{PE}(\omega_3) = \min_{t\in [0,T]} J_\mathrm{PE}\big[\op{U}_{\omega_3,u(t)}(t)\big],
\end{equation}
where $J_\mathrm{PE}$ is the perfect entangler functional, see \emref{} for details, 
and $\op{U}_{\omega_3,u(t)}(t)$ denotes the time evolution operator. 
To isolate the impact of coherent errors, we neglect decoherence due to system-environment interactions and assume purely unitary dynamics. The evolution 
$\op{U}_{\omega_3,u(t)}(t)$ is determined by both the spectator frequency $\omega_3$ and the control pulse $u(t)$. Evaluating Eq.~\eqref{eq:J} as a function of the spectator frequency $\omega_3$ yields a spectrum in which peaks indicate crosstalk due to the presence of the (third) spectator qubit~\cite{Krauss25}. 
The quantum optimal control toolkit comprises gradient-based as well as gradient-free optimization methods~\cite{KochEPJQT22}.
In gradient-based optimal control,  $\mathcal{J}_\mathrm{PE}(\omega_3)$ is extremized, which 
yields a set of coupled equations, including an update rule for the control $u(t)$.  Alternatively, a gradient-free optimization approach directly evaluates $\mathcal{J}_\mathrm{PE}(\omega_3)$ to update the control $u(t)$. In both approaches, the optimization is carried out iteratively (for a fixed frequency $\omega_3$); the details are described in \emref.

%For common tunable coupler architectures, typically only two logical qubits are coupled with a tunable coupler, but extensions to more qubits have been suggested~\cite{Huber24}.

%==================================================
\paragraph{Tunable coupler architecture}
%==================================================
To illustrate crosstalk mitigation via the perfect entangler spectrum, we consider a system of transmon qubits interacting through a tunable coupler. A minimal model comprises two computational qubits, which are the targets of the gate operation, and a single spectator qubit. All three transmons are coupled to a common flux-tunable element~\cite{McKayPRA16},
\begin{eqnarray}
  \op{H}&=& \omega^{\mathrm{max}}_c u(t) \op{b}^\dagger\op{b}
      - \frac{\alpha_c}{2} \op{b}^\dagger\op{b}^\dagger\op{b}\op{b}
      +\sum^3_{j=1} \bigg(\omega_j \op{a}_j^\dagger\op{a}_j
  \nonumber\\*
    & &\qquad
      - \frac{\alpha_j}{2}\op{a}_j^\dagger\op{a}_j^\dagger\op{a}_j\op{a}_j
    + g_j \big(\op{b}+\op{b}^\dagger)\big(\op{a}_j + \op{a}_j^\dagger\big) \bigg)\,.
  \label{eq:tunable-coupler-hamiltonian}
\end{eqnarray}
Here, $\omega^{\mathrm{max}}_c$ and $\alpha_c$ represent the maximal frequency and anharmonicity of the tunable coupler, while 
$\omega_j$, $\alpha_j$ and $g_j$ denote the frequencies, anharmonicities, and coupler-interaction strengths of the qubits.
The system is controlled with a time-dependent modulation $u(t)$ of the coupler frequency,  
\begin{subequations}
  \label{eq:time-dependent-pulse}
  \begin{equation}
    \label{eq:time-dependent-freq}
    u(t) = \sqrt{\big|\cos{\big(\pi\,\Phi(t)\big)}\big|}\,,
  \end{equation}
  where
  \begin{equation}
    \label{eq:time-dependent-flux}
    \Phi(t) = \Theta + \delta\cos(\omega_\phi t + \varphi)\,
  \end{equation}
\end{subequations}
is the magnetic flux with 
%The parameters of the flux consist of the 
amplitude $\delta$,  offset $\Theta$, phase $\varphi$ and frequency $\omega_\phi$.
In this setup, gates are implemented using a flux frequency $\omega_\phi$ that matches certain transition frequencies of the qubits~\cite{McKayPRA16,ReagorSA18,CaldwellPRA18,GanzhornPRR20,PetrescuPRA23}. For example,
the \sqrtiswap{} gate is realized by driving the transition $\ket{01}\leftrightarrow\ket{10}$ of the two targeted qubits~\cite{McKayPRA16}, whereas for the CZ gate the state $\ket{11}$ acquires a phase from a transition to the state $\ket{02}$ and back~\cite{ReagorSA18}.
All other parameters need to be fine-tuned to implement the desired gate. 
%During gate execution, the tunable coupler is supposed to remain in its ground state.
%For the calculation of the PE-spectrum, all gates are implemented between the first two qubits, while the third qubit is used as spectator qubit.

For the tunable coupler model~\eqref{eq:tunable-coupler-hamiltonian}, two types of resonances may cause dynamic crosstalk, and the PE spectrum allows one to diagnose which mechanism is relevant at a given spectator frequency~\cite{Krauss25}.
"Static" resonances occur when two or more transitions share the same frequency; they result in population exchange even without an external drive. These resonances are readily inferred from the Hamiltonian of the uncoupled transmons, without requiring knowledge of the dynamics. By contrast, drive-induced resonances are mediated by the coupling between qubits and often involve multi-photon processes. This type of crosstalk is more difficult to diagnose as it requires a full treatment of the dynamics. Irrespective of its origin, dynamic crosstalk can be suppressed by using optimal control with the PE spectrum as the cost function, as we show next.

%==================================================
%\paragraph{Optimizing for a \texorpdfstring{\sqrtiswap{}}{sqrt-iSWAP} Gate}
\paragraph{Gradient-based optimal control}
%==================================================
\begin{figure*}[tbp]
  \includegraphics[width=\textwidth]{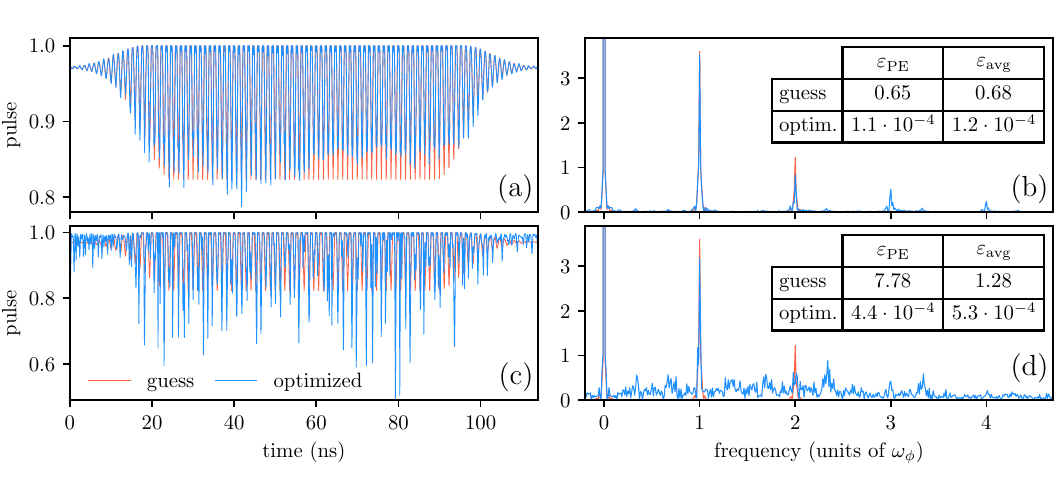}
  \caption{
    Pulse for the original \sqrtiswap{} protocol from Ref.~\cite{McKayPRA16} (red) and optimized pulses (blue) for $\omega_3=4.464\,\mathrm{GHz}$ (a,b) and $\omega_3=5.568\,\mathrm{GHz}$ (c,d). 
    The tables list the optimization error $\varepsilon_\mathrm{PE}$ and the average gate error $\varepsilon_\mathrm{avg}$.
  }
  \label{fig:mckay-oct}
\end{figure*}
We start by minimizing Eq.~\eqref{eq:J} using Krotov's method which is a powerful technique with excellent convergence properties capable of yielding high-fidelity solutions even for difficult optimization problems~\cite{ReichJCP12,GoerznQI17,PatschPRA18,GoerzSP19}. The iterative searches are initialized using the pulses derived analytically in Refs.~\cite{McKayPRA16,ReagorSA18}, where spectator qubits were not considered, 
for the \sqrtiswap{} and CZ gates. We focus below on spectator frequencies where the gates are plagued by crosstalk, as indicated by peaks of the PE spectrum~\cite{Krauss25}.
\figureref{fig:mckay-oct} compares the original \sqrtiswap{} protocol from Ref.~\cite{McKayPRA16} with optimized protocols for two exemplary spectator frequencies, $\omega_3=4.464\,$GHz and $\omega_3=5.568\,$GHz, representing the two different origins of dynamic crosstalk. The latter corresponds to a "static" resonance at $\omega_3=\omega_1-\alpha_1$~\cite{Krauss25}. In contrast, the former is a drive-induced resonance where a three-photon process drives transitions enabled by the qubit-coupler interaction, resulting in an excitation swap between the spectator qubit and the tunable coupler \cite{Krauss25}. The optimized protocols achieve comparable performance both in terms of the error $\varepsilon_\mathrm{PE}$ corresponding to Eq.~\eqref{eq:J}, which represents the optimization target, and the average gate error $\varepsilon_\mathrm{avg}$. In both cases, initial errors of order unity are reduced to values well below $10^{-3}$ as shown in Fig.~\ref{fig:mckay-oct}. 

The performance gain in these two examples involves different levels of pulse modification. The protocol in \figref[(a,b)]{fig:mckay-oct} requires only minor adaptations of the initial guess pulse, consisting of a few higher harmonics of the original driving frequency $\omega_\phi$.
In contrast, the pulse depicted in \figref[(c,d)]{fig:mckay-oct} undergoes more drastic changes, and its spectrum does not offer an immediate interpretation. 
Remarkably, the example in \figref[(a,b)]{fig:mckay-oct} is not an exception. We find for certain spectator frequencies $\omega_3$ that adjusting only a few frequency components is sufficient to mitigate crosstalk. This suggests that in such cases, a gradient-free optimization involving only the parameters $\Theta$, $\delta$, $\omega_\phi$ and $\varphi$ from \cref{eq:time-dependent-flux} may be sufficient to effectively suppress the unwanted interactions. An approach involving such minimal modifications of the protocol would be advantageous in terms of both experimental feasibility of the pulse shaping and ease of calibration.

%==================================================
\paragraph{Parameter optimization}
%==================================================

\begin{figure*}[tbp]
  \includegraphics[width=\textwidth]{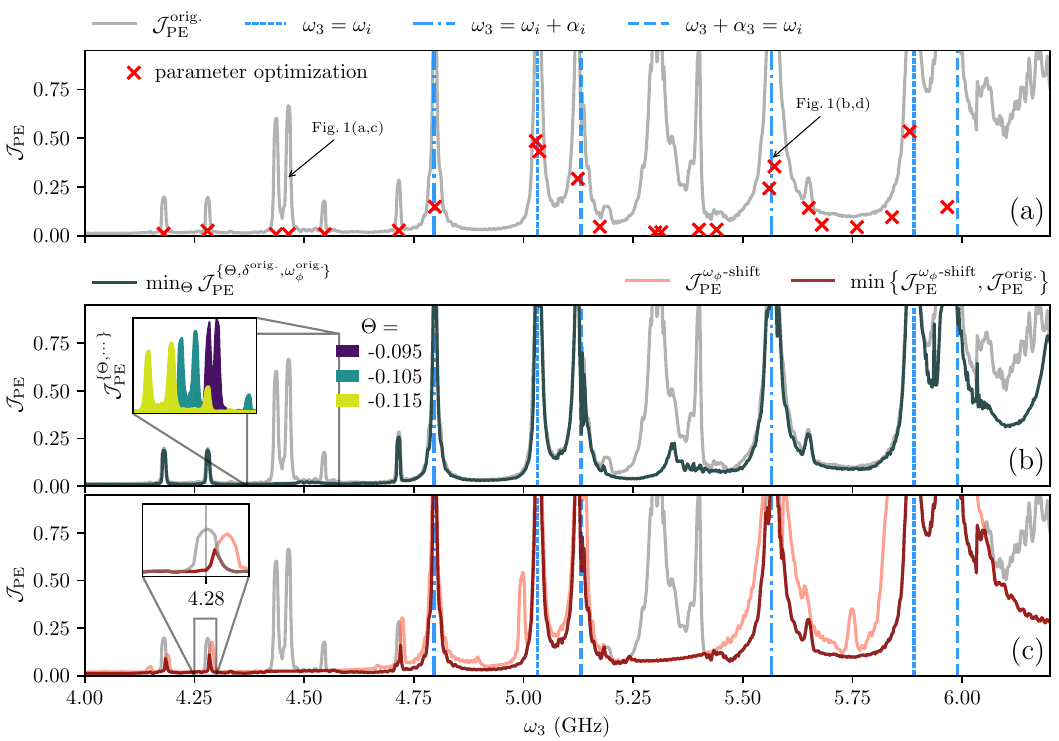}
  \caption{
    (a) Optimization errors (red crosses) obtained with gradient-free optimization for different values of $\omega_3$ for the \sqrtiswap{} gate;  the PE spectrum of the original protocol~\cite{McKayPRA16} %$\mathcal{J}_\mathrm{PE}^{\mathrm{orig.}}$ (gray). 
    is shown in gray.
    Vertical blue lines indicate "static" resonances where a transition in the spectator qubit matches a transition in one of the targeted qubits.
    (b) Optimized PE spectrum (dark green) with the pulse offset $\Theta$ as the only optimization parameter.
    %identified as strategy for the optimization of $\omega_3=4.436\mathrm{GHz}$. 
    The inset depicts the shift of the three peaks near $\omega_3=4.436\mathrm{GHz}$ as $\Theta$ is varied. %The dark green line shows the minimum of the PE-spectrum from a variety of different offsets $\Theta$.
    (c) Optimized PE spectrum $\mathcal{J}_\mathrm{PE}^{\omega_\phi\text{-shift}}$ (light red) when adjusting only $\omega_\phi$.
    %-shifted strategy obtained for $\omega_3=4.280\mathrm{GHz}$ (marked in inset). 
    The dark red curve shows the minimum between $\mathcal{J}_\mathrm{PE}^{\omega_\phi\text{-shift}}$ and the original spectrum (gray).
  }
  \label{fig:mckay-optimized}
\end{figure*}
\figureref[(a)]{fig:mckay-optimized} summarizes a series of parameter optimizations (see \emref), where Eq.~\eqref{eq:J} was minimized at spectator frequencies $\omega_3$, corresponding to peaks in the original PE spectrum  (shown in grey).   The figure demonstrates that parameter optimization effectively eliminates several of these peaks, while others are partially diminished. The latter arise from "static" resonances (blue vertical lines in \figref{fig:mckay-optimized}), whereas the eliminated peaks are all due to drive-induced resonances~\cite{Krauss25}.
This distinction arises because "static" resonances occur when two transition energies are degenerate, preventing the drive from addressing them independently. Although optimized pulses can mitigate such resonances to some degree, doing so necessitates the complex dynamics and pulse shapes illustrated in \figref[(b,d)]{fig:mckay-oct} which lack a straightforward physical interpretation. In addition to the peaks due to "static" or dynamic resonances, a distinct feature occurs in the spectral range near $\omega_3=5.75\,\mathrm{GHz}$ which is characterized by the overlapping tails of two broad peaks. In this region, crosstalk originates from a variation in the transition frequency $\Delta\omega_{12}=\omega_2-\omega_1$ that depends on the state of the spectator qubit. 
This dependency leads to asynchronous gate generation between the ground and excited spectator subspaces, a condition that is particularly difficult to mitigate. Beyond this specific region, all other spectral peaks either correspond to "static" resonances or involve crosstalk that can be suppressed using the two straightforward strategies discussed below.
%==================================================
\paragraph{Strategies for mitigating crosstalk}
%==================================================

Examining the successful parameter optimizations in \figref[(a)]{fig:mckay-optimized} in more detail, %while the specific optimal values vary, 
two main strategies emerge: shifting the offset $\Theta$ and adjusting the frequency $\omega_\phi$, cf. \cref{eq:time-dependent-flux}.
The role of the offset shift is explored in \figref[(b)]{fig:mckay-optimized}, where $\Theta$ is varied while all other parameters remain fixed.
This approach is particularly effective for the three spectral peaks near $\omega_3=4.5\,\mathrm{GHz}$.
%These correspond to unwanted transitions between the logical qubits and the tunable bus, which leads to crosstalk.
At these spectator frequencies, the drive induces an excitation exchange between the targeted qubits and the tunable coupler, leading to undesired population of the latter. Shifting the offset $\Theta$ modifies the average coupler frequency $\bar\omega_c = \omega_c^\mathrm{max}\bar{u}$, thereby altering the resonance condition responsible for the crosstalk. The shift is reflected in the PE spectrum where the three peaks move with $\Theta$, as illustrated in the inset of \figref[(b)]{fig:mckay-optimized}.
Consequently, crosstalk arising from a spectator frequency in this region is readily mitigated by an appropriate choice of $\Theta$.
To show that this strategy is consistently effective in this region, we recorded multiple spectra for various values of $\Theta$ and selected the minimum PE value for each $\omega_3$.
The resulting envelope, plotted as dark green line in \figref[(b)]{fig:mckay-optimized}, confirms the successful suppression of the triple peak near $\omega_3=4.5\,\mathrm{GHz}$. Furthermore, shifting $\Theta$ significantly flattens the peak near $\omega_3=5.3\,\mathrm{GHz}$ which also originates from an unwanted population swap between the tunable coupler and a targeted qubit. While a simple $\Theta$-shift does not eliminate this peak entirely, further adjustment of the remaining parameters achieves full mitigation, as demonstrated by the optimization results in \figref[(a)]{fig:mckay-optimized}.

The second strategy, relevant in particular  near $4.18\,\mathrm{GHz}$, $4.28\,\mathrm{GHz}$ and $4.72\,\mathrm{GHz}$, is associated with adjusting the drive frequency $\omega_\phi$. It is explored further in 
\figref[(c)]{fig:mckay-optimized} where the bright red line displays the PE spectrum $\mathcal{J}_\mathrm{PE}^{\omega_\phi\text{-shift}}$ obtained by shifting $\omega_\phi$ while using  the results of the full parameter optimization at $\omega_3=4.28\,\mathrm{GHz}$ for all other parameters. %, for which the shift of $\omega_\phi$ was particularly pronounced. 
%However, it is not sufficient to only shift the value of $\omega_\phi$, but all other parameters need to be adjusted as well in order to obtain a good gate performance.
Additionally, the dark red line depicts the minimum between $\mathcal{J}_\mathrm{PE}^{\omega_\phi\text{-shift}}$ and the original spectrum, $\mathcal{J}_\mathrm{PE}^{\mathrm{orig.}}$, shown in gray. Analysis of the peak near $\omega_3=4.28\,\mathrm{GHz}$ in the inset of \figref[(c)]{fig:mckay-optimized} reveals that adjusting $\omega_\phi$ %does not eliminate the crosstalk peaks but rather shifts them 
shifts the crosstalk peaks to other spectator frequencies $\omega_3$. This is readily explained by the fact that these peaks originate from an unwanted transition between the spectator qubit and one of the logical qubits, with transition frequencies $\omega_\phi$ and $2\omega_\phi$, respectively~\cite{Krauss25}. Thus, a change in $\omega_\phi$ results in the corresponding resonances occurring at different spectator frequencies $\omega_3$.
The strategy of shifting $\omega_\phi$ mitigates similar crosstalk peaks as shifting the offset $\Theta$ but the latter approach is more effective, eliminating some of the crosstalk peaks in \figref[(b)]{fig:mckay-optimized} completely. In contrast, peak shifts due to a modified $\omega_\phi$ may still result in non-zero overlap with the original spectrum, as seen in the inset of \figref[(c)]{fig:mckay-optimized}.
The reason why both strategies address the same crosstalk peaks is that both effectively  change the average coupler frequency compared to the reference protocol. 
%In contrast to the first protocol, this shift is not caused by a simple shift in $\Theta$, but rather, it is an effect of the modification of all parameters. This gives rise to a more pronounced offset-shift, leading to a significant displacement of all those peaks, which were mitigated by the first strategy.

%==================================================
\paragraph{CZ gate}
%==================================================

\begin{figure*}[tbp]
  \includegraphics[width=\textwidth]{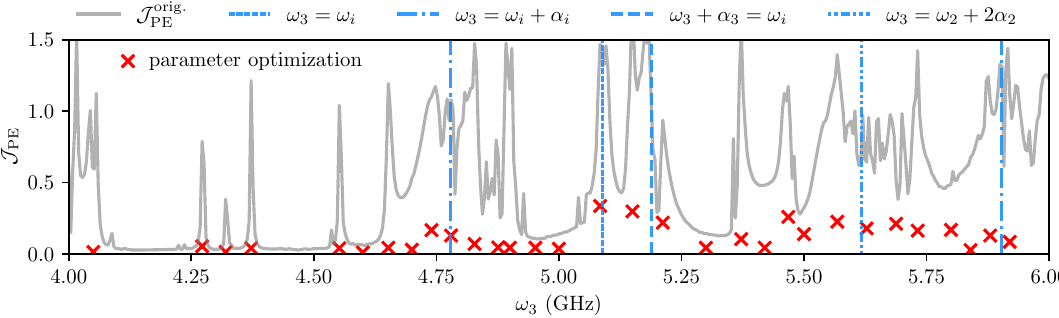}
  \caption{
  Same as \figref[a]{fig:mckay-optimized} but for the CZ gate with the original protocol found in Ref.~\cite{GanzhornPRR20}.
  }
  \label{fig:ganzhorn-optimized}
\end{figure*}
Mitigating crosstalk via pulse shaping is not restricted to a specific entangling gate protocol. \figureref{fig:ganzhorn-optimized} presents the optimized PE spectrum for a CZ protocol, where the results closely mirror those of the \sqrtiswap{} gate. While interactions associated with "static" resonances are more difficult to suppress, crosstalk arising from drive-induced resonances can be significantly mitigated and, in some cases, eliminated altogether. In particular, all crosstalk peaks at spectator frequencies $\omega_3 < 4.7\,\mathrm{GHz}$ are consistently suppressed, with analysis revealing control strategies similar to those discussed above.

In contrast, the region near $5.6\,\mathrm{GHz}$ exhibits persistent crosstalk.  In addition to the "static" resonance indicated by the blue line, dressing by the spectator qubit shifts the $\ket{11}\leftrightarrow\ket{02}$ gate transition frequency. Furthermore, dynamic resonances involving one, two, and three photons occur in close spectral proximity, as evidenced by the multi-peak structure near $5.6\,\mathrm{GHz}$~\cite{Krauss25}. This interplay of several crosstalk mechanisms impedes optimization success when only a few parameters are varied.
The spectral region $5.25\,\mathrm{GHz} < \omega_3 < 5.5\,\mathrm{GHz}$ likewise poses challenges to the optimization. Overall, the straightforward strategy of adjusting only the parameters of Eq.~\eqref{eq:time-dependent-flux}, instead of full pulse shaping (as in Fig.~\ref{fig:mckay-oct}), is capable of mitigating crosstalk in between the peaks of the original PE spectrum. 
While the denser PE spectrum of the CZ gate is more challenging to optimize than that of the
\sqrtiswap{} gate, effective mitigation remains possible across defined spectator frequency regions.

%==================================================
\paragraph{Conclusions}
%==================================================

We have demonstrated how quantum optimal control can be employed to mitigate unintended interactions driven by the control fields themselves, via suitable pulse shaping. Our approach utilizes the perfect entangler (PE) spectrum to define a loss function that directly targets crosstalk suppression. When integrated into gradient-based optimization algorithms, this method consistently improves both crosstalk errors and average gate fidelities by more than three orders of magnitude across a broad spectator frequency range. The complexity of the resulting pulse shapes reflects the underlying crosstalk mechanism. Specifically, mitigating crosstalk due to degenerate transition frequencies, which are directly apparent from the bare qubit Hamiltonian, requires highly complex pulses. This suggests that such resonances are more effectively addressed through qubit design rather than pulse shaping.
In contrast, crosstalk originating from qubit-coupler interactions, which often involve multi-photon processes, requires a full treatment of the dynamics for identification but remains amenable to mitigation through minor modifications of parametric protocols. 

Our approach complements existing strategies for suppressing crosstalk in single-qubit gates~\cite{ZhouPRL23} and can be integrated with pulse and device characterization protocols, as already established at the level of single-qubit operations~\cite{GenoisPRAppl2025}. By providing a unified control framework to identify the physical limits of crosstalk suppression, our work provides a firm basis for the development of high-fidelity, scalable quantum hardware. 
Extending such PE-based optimization to alternative architectures, such as ion traps~\cite{Strohm24}, offers a direct way to identify and mitigate platform-specific crosstalk mechanisms. Ultimately, this approach enables the development of robust control strategies across the full spectrum of current and future quantum computing platforms.

%##################################################
\begin{acknowledgments}
  Funding from the Deutsche Forschungsgemeinschaft (DFG, German Research
  Foundation)–Projektnummer 277101999–TRR 183 (project C05) and from the German
  Federal Ministry of Education and Research (BMBF) within the project QCStack
  (13N15929) is gratefully acknowledged.
\end{acknowledgments}

%##################################################
%##################################################
%##################################################

\bibliography{references}

%\clearpage

\appendix
%\begin{widetext}
    \section{--- End Matter ---} % not sure how to typeset this
\label{sec:end-matter}
%\end{widetext}
\begin{table}[tbp]
    \caption{Parameters for \sqrtiswap{} gate~\cite{McKayPRA16} and CZ gate~\cite{GanzhornPRR20}.}
    \label{tab:parameters-table}
    \begin{ruledtabular}
      \begin{tabular}{llcrr}
        & & & \sqrtiswap{} & CZ \\
      \colrule
        \multirow{3}{4em}{Qubit 1}
          & Frequency     & $\omega_1/2\pi$ & $5.8899\,\mathrm{GHz}$ & $5.089\,\mathrm{GHz}$ \\
          & Anharm.       & $\alpha_1/2\pi$ & $324\,\mathrm{MHz}$    & $310\,\mathrm{MHz}$   \\
          & Coupling      & $g_{1}/2\pi$    & $100.0\,\mathrm{MHz}$  & $116\,\mathrm{MHz}$   \\
      \colrule
        \multirow{3}{4em}{Qubit 2}
          & Frequency     & $\omega_2/2\pi$ & $5.0311\,\mathrm{GHz}$ & $6.189\,\mathrm{GHz}$ \\
          & Anharm.       & $\alpha_2/2\pi$ & $235\,\mathrm{MHz}$    & $286\,\mathrm{MHz}$   \\
          & Coupling      & $g_{2}/2\pi$    & $71.4\,\mathrm{MHz}$   & $142\,\mathrm{MHz}$   \\
      \colrule
        \multirow{3}{4em}{Qubit 3}
          & Frequency     & $\omega_3/2\pi$ &  varied               & varied                \\
          & Anharm.       & $\alpha_3/2\pi$ &  $100\,\mathrm{MHz}$  & $100\,\mathrm{MHz}$   \\
          & Coupling      & $g_{3}/2\pi$    &  $85.0\,\mathrm{MHz}$ & $85\,\mathrm{MHz}$    \\
      \colrule
        \multirow{2}{4em}{Tunable coupler}
          & Frequency     & $\omega^{\mathrm{max}}_c/2\pi$ & $7.445\,\mathrm{GHz}$ & $8.1\,\mathrm{GHz}$ \\
          & Anharm.       & $\alpha_c/2\pi$                & $230\,\mathrm{MHz}$   & $235\,\mathrm{MHz}$ \\
      \colrule
        \multirow{4}{4em}{Guess Drive}
          & Offset        & $\Theta$           & $-0.108$              & $0.15$                 \\
          & Amplitude     & $\delta$           & $0.155$               & $0.19$                 \\
          & Frequency     & $\omega_\phi/2\pi$ & $850.6\,\mathrm{MHz}$ & $816.58\,\mathrm{MHz}$ \\
          & Pulse flank   & $\sigma_t$         & $8.3\,\mathrm{ns}$    & $13\,\mathrm{ns}$      \\
      \end{tabular}
    \end{ruledtabular}
  \end{table}
The parameters used for the tunable coupler setup and flux modulation to implement the \sqrtiswap{} and CZ gates~\cite{McKayPRA16,GanzhornPRR20} are recalled in Table~\ref{tab:parameters-table}. 

%==================================================
\paragraph{The generalized perfect entangler functional}
%==================================================

The PE spectrum~\cite{Krauss25}, Eq.~\eqref{eq:J}, is based on a generalization of the PE functional for two qubits which reads~\cite{WattsPRA15,GoerzPRA15}
\begin{equation}
    J^{(2q)}_{\mathrm{PE}}[\op{U}] =
      \frac{1}{5}\big(g_3\sqrt{g_1^2+g_2^2}-g_1\big)
      + \frac{4}{5} \Delta_U\,.
    \label{eq:perfect-entangler-functional}
\end{equation}
It determines the proximity of a two-qubit gate $\op{U}$ to the set of all perfect entanglers in terms of the Euclidean distance of the local invariants $g_i$~\cite{MakhlinQIP02,ZhangPRA03}; $\Delta_U=\left(1-\Tr\left[\op{U}^\dagger\op{U}\right]/4\right)$ measures unitarity of $\op U$ in the computational subspace.
This definition can be extended to three qubits by evaluating the two-qubit PE  functional for the subspaces with the third qubit being in its ground and excited state, respectively, 
\begin{equation}
  J_\mathrm{PE}[\op{U}] = \left(
      J^{(2q)}_{\mathrm{PE}}\left[\op U^{(\ket{0})}\right]
      + J^{(2q)}_{\mathrm{PE}}\left[\op U^{(\ket{1})}\right]
      + \frac{\mathcal{S}[\op{U}]}{2}
    \right)\,.
  \label{eq:full-functional}
\end{equation}
Here $\op U^{(\ket{i})}$ is the two-qubit gates in the subspace of the third qubit being in state $\ket{i}$, and  
\begin{equation}
  \mathcal{S}[\op{U}] = 1- \bigg|
      \!\Tr\Big[
        \big(U^{(\ket{0})}\big)^\dagger U^{(\ket{1})}
      \Big]/4
    \bigg|^2 \,,
  \label{eq:similarity-functional}
\end{equation}
measures  the similarity between the gates in the two subspaces~\cite{Krauss25}.
Equation~\eqref{eq:full-functional} implies that two two-qubit gate tomographies are necessary to characterize the crosstalk at a given spectator frequency~\cite{Krauss25}.

%==================================================
\paragraph{Gradient-based optimal control}
%==================================================

In order to minimize crosstalk with gradient-based optimal control, a two-stage protocol based on the variational principle at first and second order is used.
At first order, Krotov's method~\cite{ReichJCP12,GoerzSP19} is an approach to construct and evaluate the gradient with superior convergence properties. It combines the actual target at the final time $T$, $J_T[\psi(T)]$, here the perfect entangler functional, \cref{eq:full-functional}, with a time-dependent cost, 
\begin{equation}
  \label{eq:abstract-oct-func}
  J = J_T[\psi(T)]
    + \int_0^T J_t\big[u(t)\big] \,\mathrm{d}t\,,
\end{equation}
where, following Ref.~\cite{PalaoPRA2003}, we use $J_t\big[u(t)\big] = \frac{\lambda_a}{S(t)}\big[u(t)-u_\mathrm{ref}(t)\big]^2$.
The pulse $u^{(k)}(t)$ at step $k$  of the iteration is updated according to
%\begin{widetext}
  \begin{equation}
    \label{eq:krotov_update_eq}
    \Delta u^{(k)}(t) = 
        \frac{S(t)}{\lambda_a} \mathfrak{Im}\bigg\{\sum_j
          \bigg\langle \chi_j^{(k-1)}(t)
          \bigg|
          \frac{\partial\op{H}}{\partial u^{(k)}(t)}
          \bigg|
          \psi^{(k)}_j(t)
          \bigg\rangle
        \bigg\}\,,
  \end{equation}
%\end{widetext}
where $\Delta u^{(k)}(t)=u^{(k)}(t)-u^{(k-1)}(t)$.  
$\big\{\ket{\psi^{(k)}_j(t)}\big\}$ denotes the set of the computational basis states that time-evolve under the pulse $u^{(k)}(t)$.
The so-called co-states $\big\{\ket{\chi^{(k)}_j}\big\}$ are backward propagated with the boundary condition $\ket{\chi_j^{(k)}}(T) = -\nabla_{\bra{\psi_j}} J_T \Big|_{t=T}$.
The optimization is initialized with a guess pulse, taken to be the original gate protocol, and the pulse is iteratively updated until either the break condition $J_\mathrm{PE}<10^{-5}$ is met or 100 iterations are reached. In the latter case, at the second stage, the optimization is continued with a quasi-Newton method on top of GRAPE~\cite{Fouquieres2011}. Such a second-order method can further modify the pulse using the Hessian of the functional, also when the gradient becomes very small. The second stage optimization is terminated when $J_\mathrm{PE}<10^{-5}$ or 500 iterations are reached.

\paragraph{Gradient-free optimization of the PE spectrum}

For pulse shaping remaining close to Eq.~\eqref{eq:time-dependent-pulse}, 
the downhill simplex algorithm~\cite{NelderTCJ65} is a method that works well when only a few parameters are varied.
The calculations are performed within varying bounds around the original value until either a functional value $J_{\mathrm{PE}}<10^{-2}$ or a maximum of 500 (\sqrtiswap{}), respectively, 1000 (CZ) steps is reached.
%Some of the optimizations were performed within the rotating wave approximation. This approximation is typically well fulfilled, and thus the optimizations yield similar results.
For the optimizations of the $CZ$ gate, the parameter $\varphi$ was kept fixed at 0, in order to reduce the complexity of the search space and since this parameter turned out to be the least significant. In addition to optimizing the parameters of Eq.~\eqref{eq:time-dependent-pulse}, multi-parameter optimizations with 
multiples of the original frequency $\omega_\phi$, 
\begin{align}
    \Phi(t) &= \Theta + \delta_1\cos(\omega_\phi t + \varphi_1)
    \notag\\
    &\qquad\quad + \delta_2\cos(2\omega_\phi t + \varphi_2) + \delta_3\cos(3\omega_\phi t + \varphi_3)\,,
\end{align}
were carried out, 
as was suggested by the results of the gradient-based optimization, shown in \figref[(a,c)]{fig:mckay-oct}. 
Remarkably, the results obtained for optimizations with more frequency components are similar to the single frequency results, in terms of final functional value, at the level of errors of the order of $<10^{-2}$.

\paragraph{Average gate error}
To characterize the performance of the optimized pulses, in addition to the PE functional, we also evaluate the average gate error~\cite{PedersenPLA07},
\begin{equation}
  \varepsilon_\mathrm{avg}[\op{U},\op{O}] 
    = 1-\frac{1}{20}\bigg(
        \Big|\Tr\big[\op{U}^\dagger\op{O}\big]\Big|^2
        + \Tr\big[\op{U}^\dagger\op{O}\op{O}^\dagger\op{U}\big]
    \bigg)\,.
  \label{eq:J_avg}
\end{equation}
In contrast to PE functional in \cref{eq:full-functional}, the average gate error is determined with respect to a reference gate $\op{O}$, chosen as the closest perfectly entangling unitary~\cite{WattsPRA15,GoerzPRA15,GoerznQI17}.
%To determine the gate error for a three-qubit gate, we use a similar measure as \cref{eq:full-functional}, but replace $J_\mathrm{PE}$ with $\varepsilon_\mathrm{avg}$.
%To avoid repeating the numerically demanding calculation of $\varepsilon_\mathrm{avg}$ we only calculate the average gate error for the best performing gate with respect to $J_{\mathrm{PE}}$, \ie $\op{U}_{\mathrm{min}} = \argmin_{\op{U}(t)}J_{\mathrm{PE}}[\op{U}(t)]$.

\end{document}